# Detecting Malicious PowerShell Commands using Deep Neural Networks

Danny Hendler
Ben-Gurion University
hendlerd@cs.bgu.ac.il

Shay Kels
Microsoft
shkels@microsoft.com

Amir Rubin
Ben-Gurion University
amirrub@cs.bgu.ac.il

## Abstract

Microsoft's PowerShell is a command-line shell and scripting language that is installed by default on Windows machines. Based on Microsoft's .NET framework, it includes an interface that allows programmers to access operating system services. While PowerShell can be configured by administrators for restricting access and reducing vulnerabilities, these restrictions can be bypassed. Moreover, PowerShell commands can be easily generated dynamically, executed from memory, encoded and obfuscated, thus making the logging and forensic analysis of code executed by PowerShell challenging.

For all these reasons, PowerShell is increasingly used by cybercriminals as part of their attacks' tool chain, mainly for downloading malicious contents and for lateral movement. Indeed, a recent comprehensive technical report by Symantec dedicated to PowerShell's abuse by cybercrimials [1] reported on a sharp increase in the number of malicious PowerShell samples they received and in the number of penetration tools and frameworks that use PowerShell. This highlights the urgent need of developing effective methods for detecting malicious PowerShell commands.

In this work, we address this challenge by implementing several novel detectors of malicious PowerShell commands and evaluating their performance. We implemented both "traditional" natural language processing (NLP) based detectors and detectors based on character-level convolutional neural networks (CNNs). Detectors' performance was evaluated using a large real-world dataset.

Our evaluation results show that, although our detectors (and especially the traditional NLP-based ones) individually yield high performance, an ensemble detector that combines an NLP-based classifier with a CNN-based classifier provides the best performance, since the latter classifier is able to detect malicious commands that succeed in evading the former. Our analysis of these evasive commands reveals that some obfuscation patterns automatically detected by the CNN classifier are intrinsically difficult to detect using the NLP techniques we applied.

Our detectors provide high recall values while maintaining a very low false positive rate, making us cautiously optimistic that they can be of practical value.

## 1 Introduction

Modern society is more than ever dependent on digital technology, with vital sectors such as health-care, energy, transportation and banking relying on networks of digital computers to facilitate their operations. At the same time, stakes are high for cybercriminals and hackers to penetrate computer networks for stealthily manipulating victims' data, or wreaking havoc in their files and requesting ransom payments. Protecting the ever-growing attack surface from determined and resourceful attackers requires the development of effective, innovative and disruptive defense techniques.

One of the trends in modern cyber warfare is the reliance of attackers on general-purpose software tools that already preexist at the attacked machine. Microsoft PowerShell[1] is a command-line shell and scripting language that, due to its flexibility, powerful constructs and ability to execute scripts directly from the command-line, became a tool of choice for many attackers. Several open-source frameworks, such as PowerShell Empire[2] and PowerSploit[3] have been developed with the purpose of facilitating post-exploitation cyber-offence usage of PowerShell scripting.

While some work has been done on detecting malicious scripts such as JavaScript [2, 3, 4, 5], PowerShell, despite its prominent status in the cyber warfare, is relatively untreated by the academic community. Most of the work on PowerShell is done by security practitioners at companies such as Symantec [1] and Palo Alto Networks[6]. These publications focus mainly on surveying the PowerShell threat, rather than on developing and evaluating approaches for detecting malicious PowerShell activities. The discrepancy between the lack of research on automatic detection of malicious PowerShell commands and the high prevalence of PowerShell-based malicious cyber activities highlights the urgent need of developing effective methods for detecting this type of attacks.

Recent scientific achievements in machine learning in general, and deep learning [7] in particular, provide many opportunities for developing new state-of-the-art methods for effective cyber defense. Since PowerShell scripts contain textual data, it is natural to consider their analysis using various methods developed within the Natural Language Processing (NLP) community. Indeed, NLP techniques were applied for the sentiment analysis problem [8], as well as for the problem of detecting malicious non-PowerShell scripts [5]. However, adapting NLP techniques for detecting malicious scripts is not straightforward, since cyber attackers deliberately obfuscate their script commands for evading detection [1].

In the context of NLP sentiment analysis, deep learning methods considering text as a stream of characters have gained recent popularity and have been shown to outperform state of art methods [9, 10]. To the best of our knowledge, our work is the first to present an ML-based (and, more specifically, deep-learning based) detector of malicious PowerShell commands. Motivated by recent successes of character-level deep learning methods for NLP, we too take this approach, which is compelling in view of existing and future obfuscation attempts by attackers that may foil extraction of high-level features.

We develop and evaluate several ML-based methods for the detection of malicious PowerShell commands. These include detectors based on novel deep learning architectures such as Convolutional Neural Networks (CNNs) [11, 12] and Recurrent Neural Networks (RNNs) [13], as well as detectors based on more traditional NLP approaches such as linear classification on top of character n-grams and bag-of-words [14].

Detecting malicious PowerShell commands within the high volume of benign PowerShell commands used by administrators and developers is challenging. We validate and evaluate our detectors using a large dataset[4] consisting of 60,098 legitimate PowerShell commands executed by users in Microsoft's corporate network and of 5,819 malicious commands executed on virtual machines deliberately infected by various types of malware, as well as of 471 malicious commands obtained by other means, contributed by Microsoft security experts.

**Contributions** The contributions of our work are two-fold. First, we address the important and yet under-researched problem of detecting malicious PowerShell commands. We present and evaluate the performance of several novel ML-based detectors and demonstrate their effectiveness on a large real-world dataset.

Secondly, we demonstrate the effectiveness of character-level deep learning techniques for the detection of malicious scripting. Our evaluation results establish that, although traditional NLP-based approaches yield high detection performance, ensemble learning that combines traditional NLP models with deep learning models further improves performance by detecting malicious commands that suc-ceed in evading traditional NLP techniques.

---

[1] https://docs.microsoft.com/en-us/powershell/
[2] https://www.powershellempire.com/
[3] https://github.com/PowerShellMafia/PowerSploit

[4] User sensitive data was anonymized.

Since the character-level deep learning approach is intrinsically language independent, we expect it can be easily adapted for detecting malicious usage of other scripting languages.

The rest of this paper is organized as follows. In Section 2, we provide background on PowerShell and how it is used as an attack vector and on some concepts required for understanding our deep-learning based detectors. In Section 3, we describe our dataset, how we pre-process commands and how our training set is constructed. A description of our detectors is provided in Section 4, followed by an evaluation of their performance in Section 5. Key related work is surveyed in Section 6. We conclude with a summary of our results and a short discussion of avenues for future work in Section 7. 8

## 2  Background

### 2.1  PowerShell

Introduced by Microsoft in 2006, PowerShell is a highly flexible system shell and scripting technology used mainly for task automation and configuration management [15]. Based on the .NET framework, it includes two components: a command-line shell and a scripting language. It provides full access to critical Windows system functions such as the Windows Management Instrumentation (WMI) and the Component Object Model (COM) objects. Also, as it is compiled using .NET, it can access .NET assemblies and DLLs, allowing it to invoke DLL/assembly functions. These built-in functionalities give PowerShell many strong capabilities such as downloading content from remote locations, executing commands directly from memory, and accessing local registry keys and scheduled tasks. A detailed technical discussion of these capabilities can be found in [16].

As typical of scripting languages, PowerShell commands can be either executed directly via the command line, or as part of a script. PowerShell's functionality is greatly extended using thousands of 'cmdlets' (command-lets), which are basically modular and reusable scripts, each with its own designated functionality. Many cmdlets are built into the language (such as the `Get-Process` and `Invoke-Command` cmdlets), but additional cmdlets can be loaded from external modules to further enrich the programmer's capabilities. The `Get-Process` cmdlet, for instance, when given a name of a machine which can be accessed in the context in which PowerShell is executed, returns the list of processes that are running on that machine. As another example, the `Invoke-Command` cmdlet executes the command provided as its input either locally or on one or more remote computers, depending on arguments. The `Invoke-Expression` cmdlet provides similar functionality but also supports evaluating and running dynamically-generated commands.

#### 2.1.1  PowerShell as an Attack Vector

While PowerShell can be configured and managed by the company IT department to restrict access and reduce vulnerabilities, these restrictions can be easily bypassed, as described by Symantec's comprehensive report about the increased use of PowerShell in attacks [1]. Furthermore, logging the code executed by PowerShell can be difficult. While logging the commands provided to PowerShell can be done by monitoring the shell that executes them, this does not necessarily provide the visibility required for detecting PowerShell-based attacks, since PowerShell commands may use external modules and/or invoke commands using dynamically-defined environment variables.

For instance, the Kovter trojan [17] uses simple, randomly generated innocent-looking environment variables in order to invoke a malicious script. One such command that appears in our dataset is "`IEX $env:iu7Gt`", which invokes a malicious script referenced by the "iu7Gt" environment variable.[5] A log of the executing shell would only show the command before its dynamic interpretation, but will not provide any data regarding the malicious script.

Although Microsoft improved the logging capabilities of PowerShell 5.0 in Windows 10 by introducing the AntiMalware Scan Interface (AMSI) generic in-terface [18], many methods of bypassing it have already been published [19, 1], thus effective forensic analysis of malicious PowerShell scripts remains challenging.

---

[5] `IEX` is an alias of `Invoke-Expression`.

In addition to the difficulty of forensic analysis, malware authors have several other good reasons for using PowerShell as part of their attacks [1]. First, since PowerShell is installed by default on all Windows machines, its strong functionality may be leveraged by cybercriminals, who often prefer using pre-installed tools for quicker development and for staying under the radar. Moreover, PowerShell is almost always whitelisted since it is benignly used by Windows system administrators [16].

Secondly, as PowerShell is able to download remote content and to execute commands directly from memory, it is a perfect tool for conducting file-less intrusions [20] in order to evade detection by conventional anti-malware tools. Finally, as we describe next, there are multiple easy ways in which PowerShell code can be obfuscated.

**PowerShell Code Obfuscation**  As described in [1], there are numerous ways of obfuscating PowerShell commands, many of which were implemented by Daniel Bohannon in 2016 and are publicly available in the "Invoke-Obfuscation" module he created [21]. Figure 1 lists a few key obfuscation methods we encountered in our data and provides examples of their usage. We now briefly explain each of them.

1. As PowerShell commands are not case-sensitive, alternating lower and upper case letters often appear in malicious commands.

2. Command flags may often be shortened to their prefixes. For instance, the "`-noprofile`" flag that excludes a PowerShell command from the execution policy can be shortened to "-nop".

3. Commands may be executed using the "`-EncodeCommand`" switch. While the design goal for this feature was to provide a way of wrapping DOS-unfriendly commands, it is often used by malicious code for obfuscation.

4. As mentioned previously, the "`Invoke-Command`" cmdlet evaluates a Power-Shell expression represented by a string and can therefore be used for executing dynamically-generated commands.

5. Characters can be represented by their ASCII values using "`[char]ASCII-VALUE`" and then concatenated to create a command or an operand.

6. Commands may be base-64-encoded and then converted back to a string using the "`FromBase64String`" method.

7. Base64 strings can be encoded/decoded in various ways (UTF8, ASCII, Unicode).

8. Yet another way of obfuscating commands is to insert characters that are disregarded by Power-Shell such as `` ` ``.

9. Command strings may be manipulated in real-time before evaluation using replacement and concatenation functions.

10. The values of environment variables can be concatenated in run-time to generate a string whose content will be executed.

11. Some malware generate environment variables with random names in every command execution.

While the ability to encode/represent commands in different ways and generate them dynamically at run-time provides for greater programming flexibility, Figure 1 illustrates that this flexibility can be easily misused. As observed by [1], "These [obfuscation] methods can be combined and applied recursively, generating scripts that are deeply obfuscated on the command line".

## 2.2 Deep Learning

In this section we provide background on deep learning concepts and architectures that is required for understanding the deep-learning based malicious PowerShell command detectors that we present in Section 4.

| ID | Description | Example |
|----|-------------|---------|
| 1 | Using alternating lower and upper case letters | `-ExecUTIONPoLICy BypASs -wiNDoWSTYLe hidDeN (NEW-objecT SYstEM.NET.wEbCLIeNt).DOWnLoADFiLE(<removed>);` |
| 2 | Using short flags | `-nop -w hidden -e <removed>` |
| 3 | Using encoded commands | `-EncodedCommand <removed>` |
| 4 | Invoke expression using its string representation | `- Invoke-Expression (("New-Object Net.WebClient")).('Downloadfile') ...` |
| 5 | Using "[char]" instead of a character | `... $cs = [char]71; $fn = $env:temp+$cs; ...` |
| 6 | Reading data in base 64 | `IEX $s=New-Object IO.MemoryStream([Convert]:: FromBase64String('<removed>'));` |
| 7 | Using UTF8 encoding | `$f=[System.Text.Encoding]::UTF8.GetString ([System.Convert]::FromBase64String(<removed>')); ...` |
| 8 | Inserting characters overlooked by PowerShell like ` | `...(new-object -ComObject wscript.shell).Popup(È-mail: <removed>@<removed>.com 'n 'nClient: <removed>"`) ...` |
| 9 | String manipulation | `... $filename.Replace('-','/') ... $env:temp + ':' + $name + '.exe ...` |
| 10 | Concatenating variables inline | `$emnuxgy='i'; $jrywuzq='x'; $unogv='e';... Invoke-Expression ($emnuxgy+$unogv+$jrywuzq+' ' ...);` |
| 11 | Using a random name for a variable in every run | `iex $env:vruuyg` |

Figure 1: Examples of PowerShell obfuscation methods.

*Artificial Neural Networks* [22, 23] are a family of machine learning models inspired by biological neural networks, composed of a collection of inter-connected artificial *neurons*, organized in *layers*. A typical ANN is composed of a single *input layer*, a single *output layer*, and one or more *hidden layers*. When the network is used for classification, outputs typically quantify class probabilities. A *Deep Neural Network* (DNN) has multiple hidden layers. There are several key DNN architectures and the following subsections provide more details on those used by our detectors.

### 2.2.1 Convolutional Neural Networks (CNNs)

A CNN is a learning architecture, traditionally used in computer vision [24, 25]. We proceed by providing a high-level description of the major components from which the CNN deep networks we use are composed.

As its name implies, the main component of a CNN is a *convolutional layer*. Assuming for simplicity that our input is a 2D grey scale image, a convolutional layer uses 2D $k \times k$ "filters" (a.k.a. "kernels"), for some integer $k$. As the filter is sliding over the 2D input matrix, the dot product between its $k \times k$ *weights* and the corresponding $k \times k$ window in the input is being computed. Intuitively, the filter slides over the input in order to search for the occurrences of some feature or pattern. Formally, given a $k \times k$ filter, for each $k \times k$ window $x$ of the input to which the filter is applied, we calculate $w^T \cdot x + b$, where $w$ is the filter's weights matrix and $b$ is a *bias* vector representing the constant term of the computed linear function. The $k^2$ weights of $w$, as well as the $k$ values of $b$, are being learnt during the training process.

Filters slide over the input in *strides*, whose size is specified in pixels. Performing the aforementioned computation for a single filter sliding over the entire input using stride $s$ results in an output of dimensions $\big((n-k)/s + 1\big) \times \big((n-k)/s + 1\big)$, called the filter's "activation map". Using $l$ filters and stacking their activation maps results in the full output of the convolutional layer, whose dimensions are $\big((n-k)/s + 1\big) \times \big((n-k)/s + 1\big) \times l$.

In order to maintain the non-linear properties of the network when using multiple convolutional layers, a *non-linear layer* (a.k.a. *activation layer*) is added between each pair of convolutional layers. The non-linear layer applies a non-linear *activation function* such as the *Rectified Linear Units* (ReLU) function $f(x) = max(0, x)$ whose properties were investigated by [26] or the hyperbolic tangent $f(x) = tanh(x)$ function.

A *max pooling* layer [27] "down-samples" neurons in order to generalize and reduce overfitting [28]. It applies a $k \times k$ window across the input and outputs the maximum value within the window, thus reducing the number of parameters by a factor of $k^2$. A *fully connected layer* connects all inputs to all outputs. Intuitively, each output neuron of the convolutional layers represents an image feature. These features are often connected to the network's outputs via one or more fully connected layers, where the weights between inputs and outputs (learnt during the training process) determine the extent to which each feature is indicative of each output class.

*Dropout layers* [29] can be used between fully connected layers in order to probabilistically reduce overfitting. Given a probability parameter $p$, at each training stage, each node in the input remains in the network with probability $p$ or is "dropped out" (and is disconnected from outputs) with probability $1 - p$. Dropout layers, as well as fully connected layers, may also appear in recurrent neural networks, described next.

### 2.2.2 Recurrent Neural Networks (RNNs)

RNNs are neural networks able to process sequences of data representing, e.g., text [30, 31], speech [32, 33, 34], handwriting [35] or video [36] in a recurrent manner, that is, by repeatedly using the input seen so far in order to process new input. We use an RNN network composed of *long short-term memory* (LSTM) blocks [37]. Each such block consists of a *cell* that stores *hidden state*, able to aggregate/summarize inputs received over an extended period of time. In addition to the cell, an LSTM block contains 3 components called *gates* that control and regulate information flow into and out of the cell. Roughly speaking, the *input gate* determines the extent to which new input is used by the cell, the *forget gate* determines the extent to which the cell retains memory, and the *output gate* controls the level to which the cell's value is used to compute the block's output.

In the context of text analysis, a common practice is to add an *embedding layer* before the LSTM layer [38, 39]. Embedding layers serve two purposes. First, they reduce the dimensionality of the input. Secondly, they represent input in a manner that retains its context. The embedding layer converts each input token (typically a word or a character, depending on the problem at hand) to a vector representation. For example, when taking a character-level approach, one can expect that the representations of all digits computed by the embedding layer will be vectors that are close to each other. When the problem benefits from a word-level representation, *word2vec* [40] embeddings represent each word as a vector such that words that share common contexts in the text corpus using which the model was trained are represented by vectors that are close to each other.

A bidirectional RNN (BRNN) network [41] is an RNN architecture in which two RNN layers are connected to the output, one reading the input in order and the other reading it in reverse order. Intuitively, this allows the output to be computed based on information from both past and future states. BRNNs have found successful applications in various fields [42, 43, 44]. For instance, in the context of the sentiment analysis problem, when processing text from the middle of a sentence, text seen in the beginning of the sentence, as well as text seen at the end the sentence, may be used by the computation.

# 3 The dataset

Our work is based on a large dataset which, after pre-processing (which we shortly describe), consists of 66,388 distinct PowerShell commands, 6,290 labeled as malicious and 60,098 labelled as clean. Malicious dataset commands belong to two types. For training and cross-validation, we use 5,819 distinct commands obtained by executing known malicious programs in a sandbox and recording all PowerShell commands executed by the program. For testing, we used 471 malicious PowerShell commands seen in the course of May 2017, contributed by Microsoft security experts. Using this latter type of malicious instances for evaluating our detection results mimics a realistic scenario, in which the detection model is trained using data generated inside a sandbox and is then applied to commands executed on regular machines.

As for clean commands, we received from Microsoft a collection of PowerShell commands executed within Microsoft's corporate network in the course of May 2017, on machines which had no indication of malware infection thirty days prior to the execution of the PowerShell command. Clean commands were split 48,094 for training and cross-validation and 12,004 for testing.

## 3.1 Pre-processing & Training Set Construction

We implemented a preprocessor whose key goals are to perform PowerShell command decoding and normalization for improved detection results. It also eliminates identical (as well as "almost identical") commands in order to reduce the probability of data leakage.

First, in order to be able to apply detection on "cleartext", our preprocessor decodes PowerShell commands that are encoded using base-64. Such commands are identified by the `-EncodedCommand` flag (or any prefix of it starting with '-e' or '-E'). All these commands undergo base-64 decoding, as otherwise they provide no useful detection data.[6]

Next, the preprocessor normalizes commands in order to reduce the probability of a data leakage problem [45] that, in our setting, may result from using almost-identical commands for training the model and for validating it. Indeed, we observed in our dataset PowerShell commands that differ only in a very small number of characters. In most cases, this was due to either the use of different IP addresses or to the use of different numbers/types of whitespace characters (e.g., spaces, tabs and newlines) in otherwise-identical commands. To avoid this problem, we replaced all numbers to asterisk signs ('*') and all contiguous sequences of whitespace characters to a single space and then eliminated duplicates.

We also observed in our dataset PowerShell *case-equivalent* commands that only differ in letter casing (see entry 1 in Figure 1). This was dealt with by ensuring that only a single command from each case-equivalence class is used for training/validation. We note that the dimensions of the dataset specified earlier relate to the numbers of distinct commands *after* this pre-processing stage.

Our dataset is very imbalanced, since the number of clean commands is an order of magnitude larger than that of malicious commands. In order to prevent model bias towards the larger class, we constructed the training set by duplicating each malicious command used for training 8 times so that the ratio of clean/malicious training commands is 1:1. We preferred to handle imbalance this way rather than by using under-sampling in order to avoid the risk of over-fitting, which may result when a neural network is trained using a small number of examples.

---

[6]Command arguments encoded in either base-64 or UTF8 (see entries 6, 7 in Table 1) are not decoded since, in these cases, the encapsulating command is available and can be analyzed by the detector.

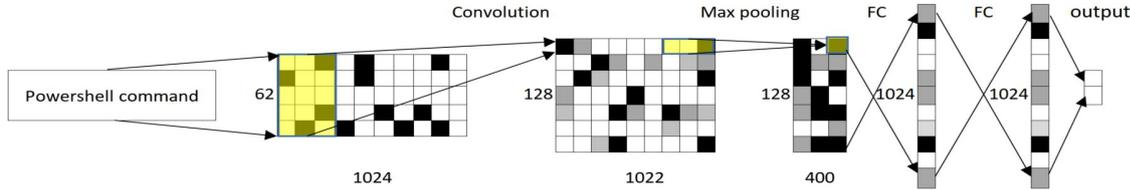

Figure 2: 4-CNN architecture used

# 4 Detection Models

In this section we describe the machine learning models we used for malicious PowerShell command detection. We then evaluate and compare their performance in Section 5.

We implemented several deep-learning based detectors. In order to assess the extent to which they are able to compete with more traditional detection approaches, we also implemented detectors that are based on traditional NLP-based methods. We proceed by describing these two sets of models.

## 4.1 Deep-Learning Based Detectors

### 4.1.1 Input Preparation

Neural networks are optimized for classification tasks where inputs are given as raw signals [24, 25]. Using these networks for text classification requires to encode the text so that the network can process it. Zhang et al. [46] explored treating text as a "raw signal at character level" and applying to it a one-dimensional CNN for text classification. We take a similar approach for classifying PowerShell commands as either malicious or benign.

First, we select which characters to encode. We do this by counting for each character the number of training set commands in which it appears and then assigning a code only to characters that appear in at least 1.4% of these commands. We have set the encoding threshold to this value because at this point there is a sharp decline in character frequency. Thus, the least-frequent character encoded (which is `) appeared in approx 1.4% of commands and the most-frequent character that was not encoded (which is a non-English character) appeared in only approx

0.3% of the training set commands.

Rare characters are not assigned a code in order to reduce dimensionality and overfitting probability. The result is a set of 61 characters, containing the space symbol, all lower-case English letters (we soon explain how we represent upper-case letters) and the following symbols: -'!%&()*,./:;?@[\]_`{|}+<=>@#$^~"

Similarly to [46], we use input feature length of 1,024, so if a command is longer than that it is truncated. This reduces network dimensions and, as shown by our evaluation in Section 5.2, suffices to provide high-quality classification. The input to the CNN network is then prepared by using "one-hot" encoding of command characters, that is, by converting each character of the (possibly truncated) command to a vector all of whose first 61 entries are 0 except for the single entry corresponding to the character's code. All characters that were not assigned a code are skipped.

In practice, we use 62-long vectors rather than 61-long vectors in order to deal with the casing of English letters. Unlike in most NLP classification tasks, in the context of PowerShell commands character casing may be a strong signal (see obfuscation method 1 in Figure 1). In order to retain casing information in our encoding, we add a "case bit", which is the 62'nd vector entry. The bit is set to 1 if the character is an upper-case English letter and is set to 0 otherwise. Thus, the representation of a PowerShell command that is being input to the CNN network is a 62x1,024 sparse matrix. A matrix representing a command that is shorter than 1,024 is padded with an appropriate number of zero columns.

As we described in Section 2.2, whereas CNNs are traditionally used for computer vision and therefore typically receive as their input a matrix representing an image, recurrent neural networks (RNNs) are optimized for processing sequences of data. Consequently, the input we provide to our RNN classifier is a vector of numbers of size at most 1,024, whose $i$'th element is the code (as described above) of the $i$'th command character (characters that were not assigned a code are skipped), except that we explicitly encode upper-case English letters since we cannot use a case bit for the RNN input representation.

### 4.1.2 Training

Stochastic gradient descent is the most widely-used method for training deep learning models [47]. We train our deep-learning based algorithms using *mini-batch gradient descent*, in which each training *epoch* (a complete pass over the training set) is sub-divided to several *mini-batches* such that the gradient is computed (and network coefficients are updated accordingly) for each mini-batch.

In order to compare all our deep-learning networks on the same basis, in all our experiments we used 16 training epochs and mini-batch size of 128. We also experimented with other numbers of epochs/mini-batches but none of them obtained significantly better classification results.

### 4.1.3 Detection models

We implemented and evaluated 3 deep-learning based detectors described in the following.

1. *A 9-layer CNN (9-CNN).* We use the network architecture designed by [46], consisting of 6 convolutional layers with stride 1, followed by 2 fully connected layers and the output layer. Two dropout layers are used between the 3 fully connected layers and a max pooling layer follows the first, second and last convolutional layers.[7] Unlike the architecture of [46] that uses fully connected layers of size 1,024 or 2,048, we use 256 entries in each such layer as this provides better performance on our data.

---

[7]Dropout and max pooling layers are typically not counted towards the network's depth.

2. *A 4-layer CNN (4-CNN).* We also implemented a shallower version of the 9-CNN architecture whose structure is depicted by Figure 2. It contains a single convolutional layer with 128 kernels of size 62x3 and stride 1, followed by a max pooling layer of size 3 with no overlap. This is followed by two fully-connected layers, both of size 1,024 – each followed by a dropout layer with probability of 0.5 (not shown in Figure 2), and an output layer.

3. *LSTM.* We implemented a recurrent neural network model composed of LSTM blocks and used the character-level representation described above. Since inputs are not sentences of a natural language, we decided not to use Word2Vec [48] embedding. Instead, our LSTM architecture contains an embedding layer of size 32. The LSTM blocks we used are bi-directional LSTM cells with output dimension of 256, followed by two fully-connected layers, both of size 256, using a dropout probability of 0.5.

## 4.2 Traditional NLP-based detectors

We used two types of NLP feature extraction methods – a character level 3-gram and a bag of words (BoW). In both we evaluated both tf and tf-idf and then applied a logistic regression classifier on extracted features. The 3-gram model performed better using tf-idf, whereas BoW performed better using tf. For each detector we selected the hyperparameters which gave the best cross-validation AUC results (evaluation results are presented in Section 5).

Note that as the 4-CNN architecture uses a kernel of length three in the first convolutional layer, the features it uses are similar to those extracted when using the character-level 3-gram detector.

## 4.3 Input Representation Considerations

Recalling the obfuscation methods used by PowerShell-base malware authors for avoiding detection (see Section 2.1.1), we observe that our input representation retains the information required for identifying them. The commands used for obfuscation, including their short versions (obfuscation method 2 in Figure 1), can be learnt due to the usage of 3-sized kernels by the deep-learning models and the usage of 3-grams by the traditional NLP models. Obfuscation method 3 is addressed by the decoding performed during data preparation (see Section 3.1).

Most other obfuscation methods (see Figure 1) use special characters such as "`", the pipe sign "|", the symbol "+" and the environment-variable sign "$". These special characters are often used when strings and the values of environment variables are concatenated in runtime for obfuscation. All these special characters appear in a significant fraction of our training set's commands and consequently they are all assigned codes by our input encoding for deep networks. They are also retained in the input provided to the traditional NLP models.

As for the usage of random names (obfuscation method 11), these typically include numbers (converted to the '*' sign) or alternating casing, and can therefore be learnt by our classifiers as well. (As we describe later, our deep learning classifiers do a better job in learning such patterns.) The usage of special strings such as "[char]", "UTF8", "Base64" or the character '`' is also covered by both models as they are retained in the input.

The only obfuscation method w.r.t. which the input to some of our detectors is superior to that provided to others is the usage of alternating lower/upper case characters (obfuscation method 1 in Figure 1). Whereas the case-bit was easily incorporated in the input to our CNN deep-learning classifiers, the RNN and the traditional NLP-based models input representations do not accommodate its usage.

## 5 Evaluation

We performed 2-fold cross validation on the training data and present the area under the ROC curve (AUC) results (rounded to the third decimal place) of our detectors in Table 1. In addition to the 5 detectors presented in Section 4, we also evaluated a variant of 4-CNN (denoted 4-CNN*) in which we did not use the case bit.

All detectors obtain very high AUC levels in the range $0.985 - 0.990$. The traditional NLP-based detectors provide excellent results in the range $0.989 - 0.990$, the 4-CNN and LSTM detectors slightly lag behind with AUC of 0.988 and 9-CNN provides a lower AUC of 0.985. The 4-CNN* detector provides slightly lower AUC than that of 4-CNN, establishing that the case bit is beneficial.

For a detector to be practical, it must not produce many false alarms. As the cyber security domain is often characterized by a very high rate of events requiring classification, even a low false-positive rate (FPR) of (say) 1% may result in too many false alarms. It is therefore important to evaluate the true positive rate (TPR) (a.k.a. *recall*) provided by detectors when their threshold is set for low FPR levels.

Table 2 presents the TPR of our detectors for FPR levels $10^{-2}$, $10^{-3}$ and $10^{-4}$ on both the training/cross-validation and the test sets. Since we have a total of about 12,000 clean commands in the test set, we stop the analysis at FPR level of $10^{-4}$. Presented values in the "Cross-validation" part of the table are the average of the two folds. Values in the "Test set" part were obtained by models trained on the training set in its entirety.

Focusing first on cross-validation results, it can be seen that, while all classifiers achieve high TPR values even for very low FPR levels, the performance of the traditional NLP detectors is better. The 3-gram detector leads in all FPR levels with a gap that increases when FPR values are decreased. Specifically, even for an FPR of 1:10,000 it provides an excellent TPR of 0.95. Among the deep-learning based detectors, 4-CNN and LSTM are superior to 4-CNN* and 9-CNN. For FPR rate of 1:10,000, 4-CNN and LSTM provide TPRs of 0.89 and 0.85, respectively. 9-CNN obtains the worst results in all experiments.

Results on the test set are significantly lower but still good. It is noteworthy that the gaps between the traditional NLP and the 4-CNN/LSTM models that we observed on the training data almost vanish on the test data. This seems to indicate that the latter models are able to generalize better.

For an FPR of 1:100, the best performers are 4-CNN and 4-CNN* with a TPR of 0.89, LSTM is second best with 0.88 and both the 3-gram and BoW

Table 1: Detectors' area under the ROC curve (AUC) values.

| 9-CNN | 4-CNN | 4-CNN* | LSTM | 3-gram | BoW |
|-------|-------|--------|------|--------|-----|
| 0.985 | 0.988 | 0.987  | 0.988 | 0.990 | 0.989 |

Table 2: TPR by FPR per model: cross-validation and test results.

| FPR | Cross-validation | | | Test set | | |
|-----|------------------|-----------|-----------|-----------|-----------|-----------|
|     | $10^{-2}$ | $10^{-3}$ | $10^{-4}$ | $10^{-2}$ | $10^{-3}$ | $10^{-4}$ |
| 9-CNN | 0.95 | 0.89 | 0.73 | 0.72 | 0.52 | 0.24 |
| 4-CNN | 0.98 | 0.96 | 0.89 | 0.89 | 0.76 | 0.65 |
| 4-CNN* | 0.97 | 0.93 | 0.85 | 0.89 | 0.72 | 0.49 |
| LSTM | 0.98 | 0.95 | 0.85 | 0.88 | 0.81 | 0.64 |
| 3-gram | 0.99 | 0.98 | 0.95 | 0.87 | 0.83 | 0.66 |
| BoW | 0.98 | 0.93 | 0.87 | 0.87 | 0.50 | 0.35 |

detectors obtain a TPR of 0.87. For FPR 1:1,000 the 3-gram detector is best with TPR of 0.83, only slightly better than LSTM's 0.81 TPR, and for FPR 1:10,000, all of 3-gram, 4-CNN and LSTM (ordered in decreasing performance) identify approximately two thirds of malicious commands. The significance of the case bit is evident when comparing the results of the 4-CNN and the 4-CNN* detectors on the test set for FPR level of 1:10,000. The TPR when using the case bit (4-CNN) is higher by almost one third than that when it is not used (4-CNN*). 9-CNN is the worst performer also in the test set experiments, by a wider margin than in the cross-validation tests.

As we've mentioned, the performance on the test set is significantly lower than that of cross-validation in all experiments. This is to be expected: whereas training set malicious commands were generated by running malware inside a sandbox, the malicious commands in the test set were contributed by security experts. Consequently, test set malicious commands may have been collected in different ways (e.g. by searching the Windows registry for malicious PowerShell commands) and may have been produced by malware none of whose commands are in the training set.

## 5.1 A Deep/Traditional Models Ensemble

We next show that by combining 4-CNN – our best deep learning model and 3-gram – our best traditional NLP model, we are able to obtain detection results that are better than those of each of them separately. We then analyze the type of malicious commands for which the deep model contributes to the traditional NLP one.

The *D/T Ensemble* is constructed as follows. We classify a command using both the 4-CNN and the 3-gram detectors, thus receiving two scores. If either one of the scores is 0.99 or higher, we take the maximum score, otherwise we take the average of the two scores. We evaluated the Ensemble's TPR by FPR performance on the test set in the same manner we evaluated the non-Ensemble algorithms (see Table 2). The D/T Ensemble significantly outperformed all non-Ensemble algorithms and obtained on the test set TPRs of 0.92, 0.89 and 0.72 for FPR levels of 1:100, 1:1,000 and 1:10,000, respectively.

In order to gain better visibility into the contribution of the 4-CNN detector on top of the 3-gram detector, we present in Figures 3a-3c the confusion matrixes of the 3-gram, 4-CNN and D/T Ensemble detectors on the test set. These results are obtained using the lowest threshold (for each of the algorithms) that provides an FPR of no more than $10^{-3}$. Since the test set contains approximately 12,000 clean instances, this means that the algorithms must have at most 12 false positives.

As can be seen by comparing Figures 3a and 3c, the D/T Ensemble adds 42 new detections on top of those made by the 3-gram detector, with only 4 new false positives. We analyzed these new detections in order to understand where the deep learning model is able to improve over the traditional NLP model.

Out of the new 42 detected commands, 15 commands contain a sequence of alternating digits and characters. In most cases, this sequence represented the name of the host or domain from which the command downloaded (most probably malicious) content. Recall that in our pre-processing of commands, we convert digits to asterisks (see Section 3.1), thus the host/domain name contains many asterisks in it. An example of the usage of such a name that appeared in one of the newly detected commands is: "..DownloadFile
('http://d*c*a*ci*x*.<domain>')..".

Each of these names appears only once and they are most probably generated by a domain generation algorithm (DGA) [49] used by the malware for communicating with its command and control center. Since these names are unique and seem random, the 3-gram algorithm is unable to learn their pattern, while the neural network is able to.

Figure 4a depicts an example of how such a host name is encoded in the input to the neural network. Note the pattern of alternating zeros and ones in the row corresponding to the symbol '*'. Figure 4b depicts a neural network filter of size 3 that is able to detect occurrences of this pattern. The filter contains ones in the first and third columns of the row corresponding to '*' (where the '*' symbol is expected to be found) and a zero in the middle column of that row, signifying that the character between the two digits is of no significance. When this filter is applied to the characters sequence depicted in Figure 4a, it creates a relatively strong signal. On the other hand, considering the 3-gram's feature extraction algorithm, since the character between the two digits changes from one command to the other, the model is unable to learn this pattern.

A similar argument can explain the detection of a few additional commands by the D/T Ensemble that were not detected by 3-gram. These commands contain a random sequence of characters alternating between lower and upper case, most probably generated by a DGA algorithm as well. Using the case bit provided as part of its input, 4-CNN is able to identify this pattern.

We note that in both the above cases, the PoweShell commands may include additional indications of maliciousness such as the web client or the cmdlets they use. Nevertheless, it is the ability to detect patterns that incorporate random characters and/or casing that causes 4-CNN to assign these command a score above the threshold, unlike the 3-gram detector.

Our ensemble detector had only seven false positive (FPs), which we manually inspected. Two FPs exhibited obfuscation patterns – one used `[System.Text.Encoding]::UTF8.GetString` (usage of `UTF8` was observed in 1,114 of the clean commands) and the other used the `-EncodedCommand` flag (which was observed in 1,655 of the clean commands). The remaining five FPs did not use any form of obfuscation, but they all used at least two flags such as `-NoProfile` and `-NonInteractive` (each seen in 5,014 and 5,833 of the clean commands, respectively).

## 5.2 Command Length Considerations

As previously mentioned, our detectors receive as input a 1,024-long prefix of the PowerShell command and longer commands are being truncated. As shown by our evaluation, this suffices to provide high-quality classification on our dataset.

A possible counter-measure that may be attempted by future malware authors for evading our detection approach is to construct long PowerShell commands such that malicious operations only appear after a long innocent-looking prefix consisting of harmless operations. In the following, we explain how such a hypothetic counter-measure can be thwarted.

Analyzing our dataset's distribution of command lengths, we find that the length of 86% of all malicious commands and 88% of all clean commands

**(a) 3-gram**

| actual value | p | n |
|---|---|---|
| **p'** | 373 | 98 |
| **n'** | 3 | 12001 |

**(b) 4-CNN**

| | p | n |
|---|---|---|
| **p'** | 340 | 131 |
| **n'** | 5 | 11999 |

**(c) D/T Ensemble**

| | p | n | total |
|---|---|---|---|
| **p'** | 415 | 56 | 471 |
| **n'** | 7 | 11997 | 12004 |

Figure 3: Confusion matrices for 3-gram, 4-CNN and Ensemble on test set, using thresholds resulting in FPR lower than $10^{-3}$.

(a) A sample hostname encoding (zeros removed for clarity).

(b) Filter capable of identifying alternating digits.

Figure 4: A hostname encoding and a filter which was used by the network to identify alternating digits and letters

is 1,024 or less. Moreover, the length of 96.7% of all malware commands and, more importantly, *the length of 99.6% of all clean commands is 2000 or less.* We remind the reader that all commands were used by our detectors regardless of their length – commands longer than 1,024 characters were simply truncated. Given the good performance of all detectors, we found no reason of using a longer input size. It would be straightforward to modify our detectors for accommodating inputs of size 2,048 or longer if and when the characteristics of malicious commands change such that this would be necessary. As of now, clean commands whose length exceeds 2000 are very rare, deeming them suspicious.

Figure 5 presents the command-length distributions of benign and malicious commands in our dataset for commands of length 1,024 or less. The distribution of malicious command length is relatively skewed to the right, indicating that malicious Power-Shell commands tend to be longer than benign commands. The high peak of very short malicious commands is to due to Kovter trojan commands [17] that constitute approximately 8% of the malicious commands population in our dataset.

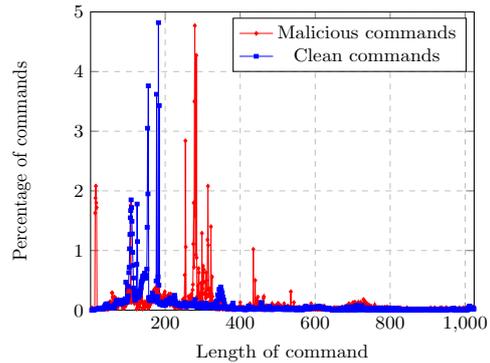

Figure 5: PowerShell command-length distributions of clean vs malicious commands.

# 6 Related work

Zhang et al. [46] introduced a deep-learning approach for text classification in which the input to convolutional neural networks (CNNs) is at character-level instead of word-level. They compared their deep-learning based classifiers with word-based traditional NLP methods (such as n-grams) and with recurrent neural networks (using LSTM blocks). Their empirical evaluation was conducted using sentiment analysis and topic classification datasets. Their results show that, while traditional methods provided better performance on small/medium datasets, character-based CNNs outperformed them on larger datasets. Our 9-CNN architecture is almost identical to theirs and its inputs are encoded in a similar manner.

Prusa and Khoshgoftaar [50] compare several architectures for short text sentiment analysis classification applied on a large dataset of tweets. They show that two relatively shallow architectures (one comprising 3 convolutional layers and 2 fully connected layers and the other comprising a single convolutional layer followed by a single LSTM layer) gave the best results. Our results are aligned with theirs in that also in our empirical evaluation the relatively shallow 4-CNN network achieved better classification performance than the deeper 9-CNN network. In both settings, classified text is relatively short – up to 140 characters inputs in their study and up to 1,024 characters in ours.

Deep learning approaches are increasingly used in recent years for malware detection. Some of these works (see [51, 52, 53, 54] for a few examples) classify programs as either malicious or benign based on their binary code and/or their runtime behaviour. In order for the neural network to be able to classify executable programs, a non-trivial feature extraction pre-processing stage is typically required whose output is fed to the neural network.

Athiwaratkun and Stokes [54] used a large dataset consisting of Windows portable executable (PE) files. They applied deep models to inputs representing the system calls made by these programs. They implemented and evaluated several models, including a character-level CNN similar to the one used by [46]. Unlike our results, in their empirical evaluation the

LSTM model achieved the best results. However, none of their neural networks was shallow.

Smith et al. also studied the problem of malware detection based on system calls made by PE executables [55]. They used several classification algorithms, including Random Forest, CNN and RNN. They observed a decay in classification quality when input length exceeded 1,000 system calls. Although problem setting and input domains differ, both our work and theirs provide evidence that limiting input length by some (domain specific) threshold may be sufficient (and is sometimes even required) for obtaining good performance.

Similarly to our work, Saxe and Berlin use deep learning models for malware detection by analyzing "cleartext" [56]. More specifically, they apply these models on a large dataset consisting of (both benign and malicious) URLs, file paths and registry keys. Their CNN architecture uses a single convolutional layer, as does our 4-CNN model.

Although some previous studies investigated the problem of detecting malicious scripting-language commands/scripts (where cleartext classification can be applied), to the best of our knowledge none of them addressed PowerShell. Several prior works presented detectors of malicious JavaScript commands by employing feature extraction pre-processing followed by the application of a shallow classifier (see, e.g., [2, 3, 4]).

Wang et al. used deep models for classifying JavaScript code collected from web pages [5]. Similarly to our work, their model uses character-level encoding, with an 8-bit character representation. They compare their classifiers with classic feature extraction based methods and study the impact of the number of hidden layers and their size on detection accuracy.

A few reports by AV vendors published in recent years surveyed and highlighted the potential abuse of PowerShell as a cyber attack vector [6, 16, 1]. Pontiroli and Martinez analyze technical aspects of malicious PowerShell code [16]. Using real-world examples, they demonstrate how PowerShell and .NET can be used by different types of malware. Quoting from their report: "Vast amounts of ready-to-use functionality make the combination of .NET and PowerShell a deadly tool in the hands of cybercriminals".

A recent comprehensive technical report by Symantec dedicated to PowerShell's abuse by cyber-crimials [1] reported on a sharp increase in the number of malicious PowerShell samples they received and in the number of penetration tools and frameworks that use PowerShell. They also describe the many ways in which PowerShell commands can be obfuscated.

Collectively, these reports shed light on the manner in which PowerShell can be used in different stages of a cyber attacks – from downloading malicious content, through reconnaissance and malware persistence, to lateral movement attempts. We have used a few of the insights they provide on PowerShell attacks for designing our detection models and for preprocessing PowerShell commands.

As we've mentioned previously, Microsoft improved the logging capabilities of PowerShell 5.0 in Windows 10, with the introduction of the AntiMalware Scan Interface (AMSI), but many methods of bypassing it have already been published. This problem was discussed and addressed in [19], where the fact that PowerShell is built on .NET architecture was used for monitoring PowerShell's activity, by leveraging .NET capabilities. As discussed in their work, the proposed solutions require some adjustments which may hurt PowerShell's performance, as well as generate some artifacts on the machine.

# 7    Discussion

PowerShell commands can be executed from memory, hence identifying malicious commands and blocking them prior to their execution is, in general, impractical. We therefore estimate that the most plausible deployment scenario of our detector would be as a post-breach tool. In such a deployment scenario, PowerShell commands that execute will be recorded and then classified by our detector. Commands classified as malicious would generate alerts that should trigger further investigation. In corporate networks, this type of alerts is typically sent to a security information and event management (SIEM) system and presented on a dashboard monitored by the organiza-tion's CISO (chief information security officer) team.

There are several ways in which this work can be extended. First, while we have implemented and evaluated several deep-learning and traditional NLP based classifiers, the design space of both types of models is very large and a more comprehensive evaluation of additional techniques and architectures may yield even better detection results.

Secondly, in this work we targeted the detection of individual PowerShell commands that are executed via the command-line. An interesting direction for future work is to devise detectors for complete Power-Shell scripts rather than individual commands. Such scripts are typically longer than single commands and their structure is richer, as they generally contain multiple commands, functions and definitions. Effective detection of malicious scripts would probably require significantly different input encoding and/or detection models than those we used in this work.

Another interesting avenue for future work is to devise detectors that leverage the information collected by Microsoft's AntiMalware Scan Interface (AMSI) [18]. As mentioned previously, AMSI is able to record PowerShell commands (generated both statically and dynamically) that are executed in run-time, so detectors may have more data to operate on. However, although AMSI may be less vulnerable to many of the obfuscation methods described in Section 2.1.1, attackers may be able to find new ways of camouflag-ing the AMSI traces of their malicious commands.

# 8    Conclusion

In this work we developed and evaluated two types of ML-based detectors of malicious PowerShell commands. Detectors based on deep learning architectures such as Convolutional Neural Networks (CNNs) and Recurrent Neural Networks (RNNs), as well as detectors based on more traditional NLP approaches such as linear classification on top of character n-grams and bag-of-words.

We evaluated our detectors using a large dataset consisting of legitimate PowerShell commands executed by users in Microsoft's corporate network, malicious commands executed on virtual machines de-liberately infected by various types of malware, and malicious commands contributed by Microsoft security experts.

Our evaluation results show that our detectors yield high performance. The best performance is provided by an ensemble detector that combines a traditional NLP-based classifier with a CNN-based classifier. Our analysis of malicious commands that are able to evade the traditional NLP-based classifier but are detected by the CNN classifier reveals that some obfuscation patterns automatically detected by the latter are intrinsically difficult to detect using traditional NLP-based classifiers. Our ensemble detector provides high recall values while maintaining a very low false positive rate and so holds the potential of being useful in practice.